\documentclass[letterpaper]{article} 
\usepackage{aaai25}  
\usepackage{times}  
\usepackage{helvet}  
\usepackage{courier}  
\usepackage[hyphens]{url}  
\usepackage{graphicx} 
\usepackage{algorithm}
\usepackage{algorithmic}
\usepackage{amsmath}
\usepackage{xcolor}
\usepackage{amsfonts} 
\usepackage{multirow}
\usepackage{booktabs}
\usepackage{natbib}  
\usepackage{caption} 
\usepackage{times}  
\usepackage{helvet}  
\usepackage{courier}  
\usepackage[hyphens]{url}  
\usepackage{graphicx} 
\usepackage{natbib}  
\usepackage{caption} 
\usepackage{algorithm}
\usepackage{algorithmic}
\usepackage{amsmath}
\usepackage{xcolor}
\usepackage{amsfonts} 
\usepackage{multirow}
\usepackage{booktabs}
\usepackage{newfloat}
\usepackage{listings}
\usepackage{algorithm}
\usepackage{algorithmic}

\usepackage{newfloat}
\usepackage{listings}
\DeclareCaptionStyle{ruled}{labelfont=normalfont,labelsep=colon,strut=off} 
\floatstyle{ruled}
\newfloat{listing}{tb}{lst}{}
\floatname{listing}{Listing}
\title{DIDiffGes: Decoupled Semi-Implicit Diffusion Models for Real-time Gesture Generation from Speech}
\author {
    Yongkang Cheng\textsuperscript{\rm 1, \rm 3},
    Shaoli Huang\textsuperscript{\rm 1}\thanks{Corresponding author: Shaoli Huang.},
    Xuelin Chen\textsuperscript{\rm 1},
    Jifeng Ning\textsuperscript{\rm 3},
    Mingming Gong\textsuperscript{\rm 2,\rm 4}
}
\affiliations {
    \textsuperscript{\rm 1}Tencent AI Lab\\
    \textsuperscript{\rm 2}School of Mathematics and Statistics, The University of Melbourne\\
    \textsuperscript{\rm 3}College of Information Engineering, Northwest A\&F University\\
    \textsuperscript{\rm 4}Mohamed bin Zayed University of Artificial Intelligence, United Arab Emirates\\
    
    cyk19990422@gmail.com,
    shaol.huang@gmail.com, 
    xuelin.chen.3d@gmail.com,
    njf@nwsuaf.edu.cn,
    mingming.gong@unimelb.edu.au
}

\usepackage{bibentry}

\begin{document}

\maketitle

\begin{abstract}
Diffusion models have demonstrated remarkable synthesis quality and  diversity in generating co-speech gestures.
However, the computationally intensive sampling steps associated with diffusion models hinder their practicality in real-world applications.
Hence, 
we present DIDiffGes, for a Decoupled Semi-Implicit Diffusion model-based framework, that can synthesize high-quality, expressive gestures from speech using only a few sampling steps.
Our approach leverages Generative Adversarial Networks (GANs) to enable large-step sampling for diffusion model. We decouple gesture data into body and hands distributions and further decompose them into marginal and conditional distributions. GANs model the marginal distribution implicitly, while L2 reconstruction loss learns the conditional distributions exciplictly.
This strategy enhances GAN training stability and ensures expressiveness of generated full-body gestures. Our framework also learns to denoise root noise conditioned on local body representation, guaranteeing stability and realism. DIDiffGes can generate gestures from speech with just 10 sampling steps, without compromising quality and expressiveness, reducing the number of sampling steps by a factor of 100 compared to existing methods. Our user study reveals that our method outperforms state-of-the-art approaches in human likeness, appropriateness, and style correctness. Project is https://cyk990422.github.io/DIDiffGes.
\end{abstract}

%

\section{Introduction}
The integration of large language models like ChatGPT~\cite{schulman2022chatgpt} into our digital lives has heralded a significant evolution in human-computer interaction. These models facilitate more natural and engaging conversations. However, to fully exploit the potential of these interactions, integrating real-time, realistic non-verbal elements, such as gestures, in harmony with the advanced verbal capabilities of these models, is crucial.

The realm of gesture synthesis faces the challenge of precisely and swiftly generating complex human gestures, encompassing hand and body movements. GAN-based methods, like those in~\cite{liu2022beat,cheng2024conditional}, although promising, struggle with motion representation and training complexities. VAE-based approaches, as seen in~\cite{yoon2020speech,yu2024signavatars}, often result in less fluid arm poses, diminishing the realism of the generated gestures. Diffusion model-based methods~\cite{cheng2024expgest,}, highlighted in studies such as~\cite{ao2023gesturediffuclip, yang2023diffusestylegesture, yang2023unifiedgesture}, have emerged as promising alternatives due to their comprehensive gesture movement capture. However, their slow generation speed is a substantial barrier, especially for real-time interactions with virtual agents.

Recent advancements like the MLD approach from~\cite{chen2023executing}, which introduces a latent space structure for encoding body movements~\cite{kingma2013auto} and adapts the latent diffusion model~\cite{rombach2022high}, along with the DDIM sampling strategy~\cite{song2020denoising}, attempt to address these challenges. Nevertheless, these methods often prioritize increased computation speed at the expense of generating overly smooth motions, thereby lacking the nuanced complexity observed in natural human movement.
\begin{figure*}[t]
  \begin{center}
    \includegraphics[width=1.0\linewidth]{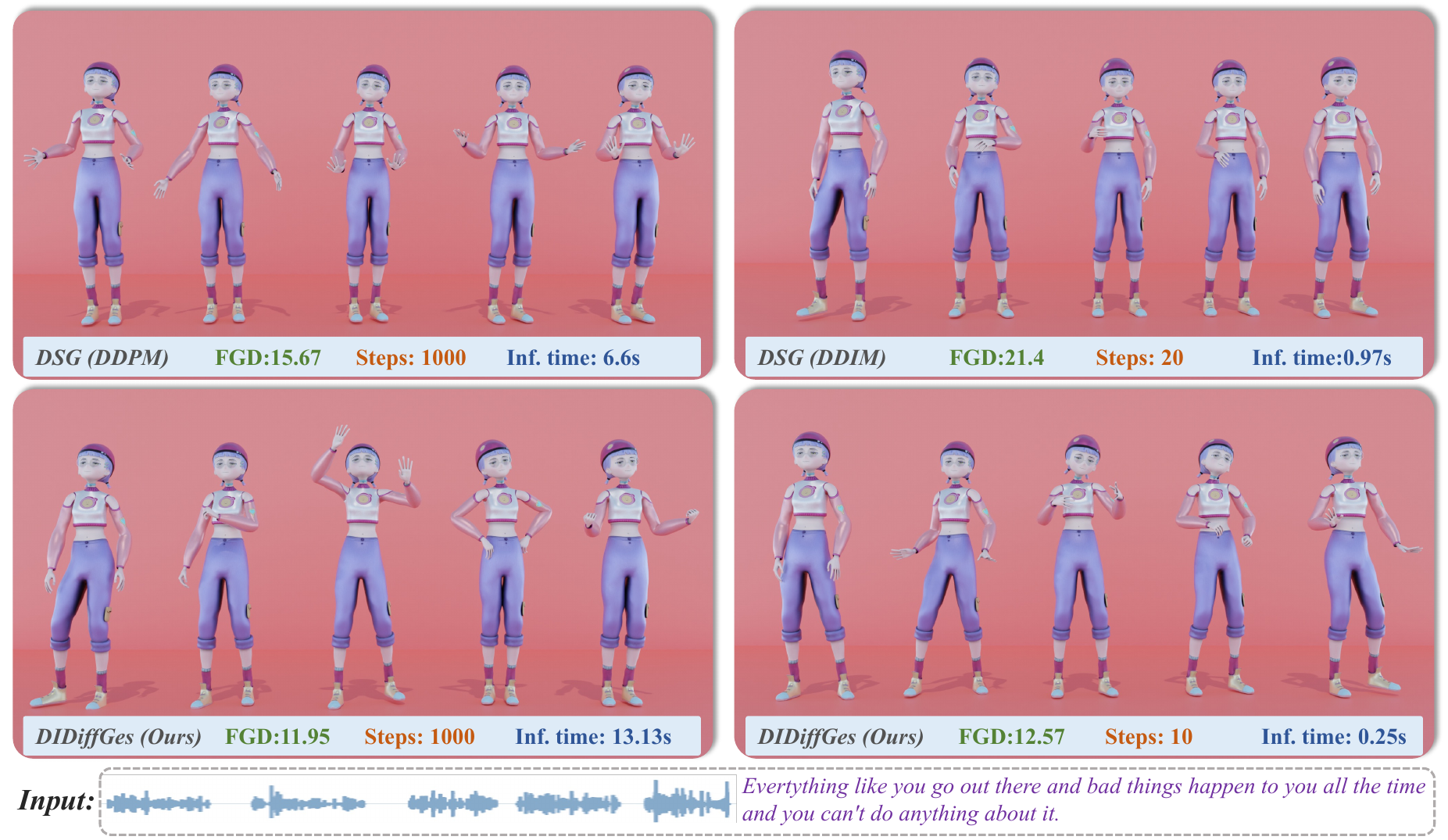}
    \vspace{-1em}
  \end{center}
  \caption{Comparison of four different sampling methods: DSG~\cite{yang2023diffusestylegesture} with DDPM, DSG with DDIM, our method with 1000-step sampling, and our method with 10-step sampling.}
  \label{fig:teaser}
  \end{figure*}
In this paper, we propose DIDiffGes, a Decoupled Semi-Implicit Diffusion model framework that aims to resolve the speed and fidelity issues in gesture generation. Our model is inspired by the recent Semi-Implicit Denoising Diffusion model (SIDDM)~\cite{xu2023semi}, which demonstrates unconditional high-fidelity image generation in a few steps. ``Semi-implicit" denotes the training of the large-step denoiser using adversarial learning as the implicit objective, complemented by L2 loss as an explicit objective for better convergence.  DIDiffGes generalizes SIDDM to human gesture generation by further decoupling the denoiser's output into distinct representations for hand and body noise, ensuring high-quality gesture generation at a few diffusion steps.

To be more specific, DIDiffGes's innovation lies in adversarially training separate noise components for body and hand movements at each diffusion step, better capturing their varying complexities and dynamics. This ensures a balanced and accurate representation of both movement types, crucial for generating realistic, expressive, and diverse gestures. The introduction of semi-implicit objectives into the diffusion framework is transformative, facilitating significant data distribution changes at each step and producing outputs closely resembling real gesture data. This capability is particularly valuable in real-time applications where fast, accurate, and natural-looking gesture generation is essential.

Moreover, DIDiffGes employ a sequential diffusion denoiser strategy that recovers root noise conditioning on local body representations. By focusing on the interplay between root motion and local body motion, this strategy improves coordination helps reduce unnatural phenomena, such as foot sliding and body drifting, which can occur during gesture generation. 

We validate our approach on both the BEATs~\cite{liu2022beat} and ZeroEGGs~\cite{ghorbani2023zeroeggs} datasets. Experimental results demonstrate that our method can generate high-quality, realistic, and natural gesture motions with fewer sampling steps. Notably, with only \textbf{10} sampling steps, our method surpasses existing open-source diffusion model-based approaches in the FGD metric and is nearly 15 times faster, inferring 88 frames of gesture motion in just 0.4 seconds. Our user study also confirms the superior quality of our rapidly generated results compared to methods requiring more sampling steps. Furthermore, we can quickly generate dance motions in the music-to-dance task, achieving performance comparable to existing methods.

In summary, DIDiffGes represents a significant advancement in the field of gesture synthesis, addressing critical needs for efficient, high-quality, and realistic gesture generation. It enhances the capabilities of large language models in real-time applications, paving the way for more natural and expressive human-AI interactions, and setting a new standard in motion synthesis across various domains.

In our open-source A2G repository~\cite{cheng2025hologest}, DIDiffGes, as an important acceleration module, helps to implement the real-time A2G generation model.

\section{Related Work}
\begin{figure*}
  \begin{center}
    \includegraphics[width=1\linewidth]{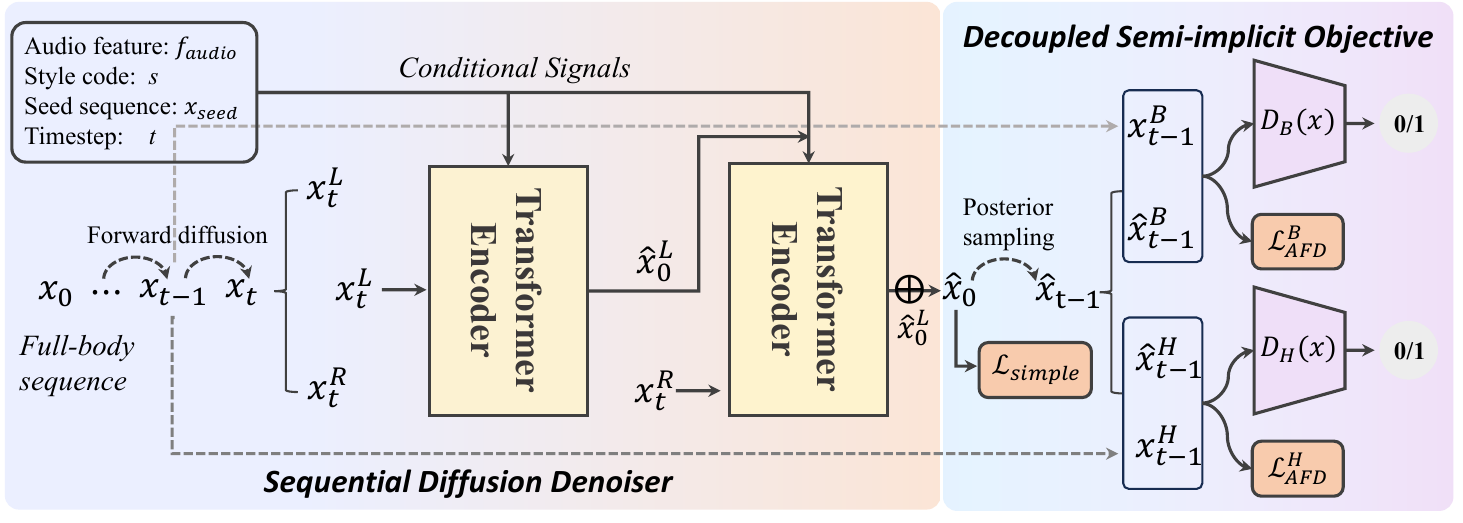}
  \end{center}
  \vspace{-5pt}
  \caption{Our learning framework integrates a Sequential Diffusion Denoiser with two transformer encoders and a Decoupled Semi-implicit Objective. The first encoder denoises local motion and provides a conditional signal for the second encoder, which denoises root noise. The final result, a combination of local motion and root result, is added with t-1 step noise via posterior sampling and then decoupled into body and hand noise. These noise undergo adversarial training against prior sampled noise, supervised by Auxiliary Forward Diffusion Loss. For a detailed description of the network architecture, please refer to our supplementary materials.}
  \vspace{-10pt}
  \label{fig:arc1}
  \end{figure*}

\textbf{Audio-to-Gesture Generation} leverages rich multimodal inputs such as speech audio, music files, and frame-missing action sequences to generate action sequences. The task of co-speech gesture generation is particularly complex, requiring understanding of speech melody, semantics, gesture action, and their interrelationships. Early data-driven methods, such as those proposed by~\cite{liu2022audio,habibie2021learning}, attempted to learn gesture matching from human demonstrations but often resulted in less diverse actions. Later works, like~\cite{habibie2021learning,yi2023generating,xie2022vector}, enhanced the model's ability to generate diverse results and introduced the concept of generating distinctive and expressive gesture results. Some studies, such as~\cite{yang2023diffusestylegesture,yang2023unifiedgesture,ahuja2020style,ao2023gesturediffuclip}, trained a unified model for multiple speakers, embedded each speaker's style in space, or introduced style transfer technology. Other works~\cite{zhou2022gesturemaster,habibie2022motion} used motion matching to generate gesture sequences, although this method often requires complex matching rules. Despite its challenges, audio-driven animation has attracted widespread attention and has been significantly improved by the recent emergence of the distinctive, high-quality dataset ZeroEGGs~\cite{ghorbani2023zeroeggs}. Our DIDiffGes stands out in key ways. It's the first in gesture gen. to use GANs for accelerating diffusion-based methods. Also, it denoises local motion noise before root noise, enhancing stability. Finally, by interfacing and separately supervising body and hands with GANs, it ensures finger movement, a unique feature.

\noindent\textbf{Diffusion Models} have shown great results in many fields~\cite{rombach2022high,chen2023executing,saharia2022palette}, esp. in human motion gen. Motion Diffuse~\cite{zhang2022motiondiffuse} was the 1st to use diffusion models for text-conditioned human motion gen., giving fine-grained instr. for 2 body parts. MDM~\cite{tevet2022human} is a key work, introducing motion diffusion model to handle the relationship between motion rep. and text control cond. Recent work has focused on motion trajectory and joint control. In our work, we capture the relationship between gesture seqs. and speech by an attention mech. to gen. highly matched results. But due to the high dim. and iterative nature of diffusion models, motion gen. based on DDPM has time overhead issues. MLD~\cite{chen2023executing} brought latent diffusion into motion gen. to enhance quality and reduce resource req., by training a VAE for motion embedding 1st and then applying latent diffusion in the latent space. However, it's a non-end-to-end method and may produce artifacts. Our DIDiffGes focuses on efficiently gen. high-quality motion. Unlike methods using DDIM or latent reps., we use GANs for large-step sampling in diffusion models.

\section{Method}
Our goal is to swiftly create high-fidelity and richly expressive co-speech gesture sequences derived from input audio signals using diffusion generative models. We ultimately aim to integrate these high-quality models into real-time applications for everyday use. To achieve this, we endeavor to expedite the denoising process by implementing a diminished number of steps and increasing step sizes. In the following sections, we elucidate our Overall Structure, Sequential Diffusion Denoiser, and Decoupled Semi-implicit Objective.

\subsection{Overall Structure}
Our framework is capable of receiving real-time audio signals of speech or pure music and guiding the generation of highly expressive full-body gestures or dance sequences. Additionally, we allow for the input of other control signals, such as style labels to enrich the emotional content of the generated gesture sequences, or seed sequences for producing smooth long-frame sequences. As illustrated in Figure~\ref{fig:arc1}, our overall architecture comprises two core components: the Sequential Diffusion Denoiser and the Decoupled Semi-implicit Objective. The former mitigates the unnaturalness of denoised motion sequences in non-physics-based environments, while the latter breaks the assumption dependency of DDPM by modeling large-stride denoising distributions to achieve fewer-step sampling, thereby enabling high-speed generation. Furthermore, decoupling the body and hand parts for independent modeling of their respective distributions significantly enhances the expressiveness of the generated sequences, providing an improved user experience. To the best of our knowledge, DIDiffGes is the first framework attempting real-time high-fidelity co-speech gesture generation, offering the possibility of practical real-time applications for current audio-to-gesture research. 

\subsection{Sequential Diffusion Denoiser}
Traditional diffusion model-based methods directly add noise to the global gesture representation, and then denoise through a simple Transformer-Based network to obtain the final generation result. However, these methods train their models in a non-physical environment, making issues such as foot sliding and global jitter inevitable. Incorporating a simulation environment during the sampling process would compromise our real-time requirements. Nevertheless, we observe that these physical sliding issues often occur in large-scale gestures. For instance, when swinging the arm, the human legs remain stationary, but our torso bends with the arm movement, often leading to incorrect root joint displacements and resulting in a sliding effect. Therefore, we further decouple the limbs, extracting the root joint's representation from the limb representation, and use the denoised local motion as a new conditional control signal to guide the generation of root motion that conforms to the local motion. This approach alleviates the physical sliding problem and enhances the visual effect without significantly increasing the time overhead, maintaining the original training stability.\\
\textbf{Preliminary. }The gesture sequence is represented as $x^{1:N}$, where $N$ denotes the number of frames in the gesture sequence. Subsequently, we employ the diffusion probability model~\cite{ho2020denoising} to generate co-speech gestures, in which the diffusion model gradually anneals pure Gaussian noise into the gesture distribution $p(x)$. As a result, the model can predict noise from the $T$-step Markov noise process $\{x_{t}^{1:N}\}_{t}^{T}$, where $x_{0}^{ 1:N}$ is directly sampled from the original data distribution. The diffusion process is as follows:
\begin{equation}
\begin{aligned}
 q(x_{t}|x_{t-1}) = \mathcal{N}(\sqrt{\frac{\alpha_{t}}{\alpha_{t-1}}}x_{t-1},(1-\frac{\alpha_{t}}{\alpha_{t-1}})I), 
\end{aligned}
\end{equation}
where $\{\beta_{t}\}_{t=1}^{T}$ is the variance schedule and $\alpha_{t} = \prod_{s=1}^{t}(1-\beta_{s})$. Subsequently, following the idea in MDM~\cite{tevet2022human}, we retain the denoising result as $\hat{x_{0}}=\epsilon_{t}^{\theta}(x_{t})$. However, due to the widespread use of the acceleration strategy DDIM, we present the description in the form of DDIM to cater to a broader audience in the diffusion model domain. To avoid further confusion, we preserve the description of $\hat{x_{0}}$ and $\epsilon$ from the DDIM paper, defining the model output as predicted noise. We rewrite the reverse process as follows:
\begin{equation}
\begin{aligned}
    \hat{x}_{0}&= \frac{x_{t}-\sqrt{1-\alpha_{t}} \epsilon_{t}^{\theta}\left(x_{t}\right)}{\sqrt{\alpha_{t}}}.\\
\end{aligned}
\end{equation}
Upon obtaining $\hat{x}_{0}$ and the known forward diffusion result $x_{t}$, we can calculate the posterior distribution and sample $\hat{x}_{t-1}$. The formula representation is as follows:
\begin{equation}
\begin{aligned}
    \hat{x}_{t-1}&=\sqrt{\alpha_{t-1}} \hat{x}_{0}+\sqrt{1-\alpha_{t-1}-\sigma_{t}^{2}} \cdot \epsilon_{t}^{\theta}\left(x_{t}\right)+\sigma_{t} z_{t}.
\end{aligned}
\end{equation}

\noindent\textbf{Decoupled Denoiser Structure. }Our structure is illustrated in Figure~\ref{fig:arc1}. We first forward-add noise to the original motion sequence $X_{0}^{1:N}$, sampled from $p(x_{0})$, to pure Gaussian noise $x_{t}^{1:N}$ using Equation (1). Subsequently, we decouple the root joint's noise representation $X_{t}^{R}$ as the noise input for the next stage, along with the audio, seed sequence, and style code, which are input into the denoiser. Our conditional denoiser contains a Transformer-based encoder and a conditional control signal encoder. The former consists of 12 layers and 8-head self-attention modules with default skip connections, while the latter is composed of Linear layers and WavLM modules. The Linear Block linearly maps the seed sequence and style code, while the audio input is encoded through WavLM to capture spectral information and is linearly mapped to the same dimensional space. We then concatenate all control signals as conditional features input into the Encoder. As shown in Equation (3) of Section 3.1, our denoiser directly predicts the clean local gesture sequence, $\hat{x}_{0}^{L}=G(x_{t},t,c)$, where $c$ is the conditional signals and $t$ is timestep. Next, we encode $\hat{x}_{0}^{L}$ as a local motion condition and additionally concatenate it to the control signals as a new guiding condition to generate the global motion sequence $\hat{x}_{0}^{R}$. Finally, we concatenate the local sequence $\hat{x}_{0}^{L}$ and global sequence $\hat{x}_{0}^{R}$ into the overall motion $\hat{x}_{0}$, and constrain it with the following reconstruction loss and physical constraints: 
\begin{equation}
\begin{aligned}
    \mathcal{L}_{simple}&=E_{x_{0}~q(x_{0}|c),t~[1,T]}[HuberLoss(x_{0}-\hat{x}_{0})],\\
    \mathcal{L}_{foot}&=\frac{1}{N-1}\sum_{i=1}^{N-1}\|\hat{x}_{0}^{i+1}-\hat{x}_{0}^{i} \|f_{i},\\
\end{aligned}
\end{equation}
where $f_{i}$ is determined by calculating the rate of change in the y-axis position of the footstep to judge whether the foot is in contact with the ground. Notably, the root joint regeneration part proposed in our method effectively optimizes the human trajectory, thereby avoiding the unnatural state of both feet being suspended simultaneously.

\subsection{Decoupled Semi-implicit Objective}
\textbf{Problem Description. }The key to implementing a real-time denoising diffusion model lies in increasing the noise addition step size to reduce the number of denoising steps, thereby achieving the goal of high-speed generation. As described in Section Introduction, DDPM is based on the assumption that the noise added at each step is small and sampled from a unimodal distribution. Hence, we can parameterize $p_{\theta}(x_{t-1}|x_{t})$ as a Gaussian Distribution, using the $L2$ loss to model the $KL$ divergence between $p_{\theta}(x_{t-1}|x_{t})$ and $q(x_{t-1}|x_{t},x_{0})$ at the same $t$. However, when we directly increase the noise step size based on this, $p_{\theta}(x_{t-1}|x_{t})$ no longer conforms to a Gaussian Distribution. Simple explicit $L2$ loss cannot model such a complex motion distribution, leading to the generation of unnatural jittery results.\\
To model complex distributions, we naturally consider the GAN model. To address this issue, some researchers~\cite{wang2022diffusion} proposes conditional generators and conditional discriminators, introducing adversarial learning strategies to model the complex motion distribution between multiple sampling steps:
\begin{equation}
\begin{aligned}
    \underset{\theta}{\min} \underset{D_{adv}}{\max} \sum\limits_{t>0} \mathbb{E}_{q(x_{t})}D_{adv}(q(x_{t-1}|x_{t})||p_{\theta}(x_{t-1}|x_{t})),
\end{aligned}
\end{equation}
Within this framework, the conditional discriminator endeavors to differentiate between the predicted denoising distribution and the original motion distribution, while the conditional generator aspires to render them indistinguishable. Nevertheless, adversarial learning constitutes a purely implicit strategy; during the training process, factors like data volume and distribution complexity may impact training stability, requiring repeated hyperparameter searches to facilitate training. When attempting to model human motion distributions with higher physical geometric constraint requirements, this purely implicit learning strategy is often proven to be statistically inefficient.\\
\textbf{Semi-implicit Matching Constraints. }Upon examining the implementation of Equation(5), it becomes evident that, during the adversarial phase, the method indirectly matches the conditional distribution by aligning with the joint distribution:
\begin{equation}
\begin{aligned}
    \underset{\theta}{\min} \underset{D_{adv}}{\max}  \mathbb{E}_{q(x_{0})q(x_{t-1}|x_{0})q(x_{t}|x_{t-1})}D_{adv}(q(x_{t-1},x_{t})|| \\p_{\theta}(x_{t-1},x_{t})),
\end{aligned}
\end{equation}
This approach requires connecting large-stride denoising distributions between two adjacent time steps in each discrimination phase of adversarial learning. However, the large-step noise distribution is often a complex multimodal distribution rather than the unimodal distribution assumed by DDPM. This undoubtedly makes GAN-based frameworks difficult to train and requires a reasonable design of the sampling step t, with the cost of searching parameters during the adversarial training phase being too high. However, adversarial training is a purely implicit matching process, typically used to constrain distributions that cannot be explicitly represented. We consider using a simpler marginal distribution to replace the joint distribution in Equation(6). That is, we directly compute the posterior distribution and subsequently perform adversarial learning with the forward process to model the large-step denoising distribution. The formula representation is as follows:
\begin{equation}
\begin{aligned}
    \underset{\theta}{\min} \underset{D_{adv}}{\max}  \mathbb{E}_{q(x_{0})q(x_{t-1}|x_{0})q(x_{t}|x_{t-1})}[-log(D_{adv}(x_{t-1},c,t))] \\
    +[-log(1-D_{adv}(\hat{x}_{t-1},c,t))],
\end{aligned}
\end{equation}
Although we have simplified the implicit matching process, making adversarial training more stable, we have encountered a new issue. Since the large-step denoising distribution is often a complex multimodal distribution, the posterior sampling $p_{\theta}(\hat{x}_{t-1}|x_{t},\hat{x}_{0})$ result still exhibits significant differences from the forward process, rendering our denoiser unable to successfully reverse from the pure noise distribution to the original distribution. Based on this, we introduce a regularization term, Auxiliary Forward Diffusion Constraint (AFD), for explicitly constraining the similarity between posterior sampling results and forward diffusion results at the same time step. Its representation is as follows:
\begin{equation}
\begin{aligned}
    \mathbb{E}_{q(x_{0})q(x_{t-1}|x_{0})q(x_{t}|x_{t-1})}\frac{(1-\beta_{t})||\hat{x}_{t-1}-x_{t-1}||^{2}}{\beta_{t}},
\end{aligned}
\end{equation}
where $\sqrt{1-\beta_{t}}x_{t-1}$ represents the mean of the forward process $q(x_{t}|x_{t-1})$, and $\beta_{t}$ represents its variance.\\
\textbf{Component Decoupling Structure.} Although our proposed semi-implicit method can better adapt to the complex human motion distribution modeling, thus achieving high-quality rapid generation, the adversarial learning strategy of GAN is characterized by learning the associations of all components within the distribution. It is evident that the motion distribution of fingers is distinct from that of limbs. Limb motion has a larger amplitude and aligns more closely with the melody, while finger motion is smaller and more precise, leaning towards semantic matching. Holistic modeling would lead GAN to fit body data more closely while neglecting finger motion, thereby reducing the expressiveness of the overall gesture. Based on this, we decouple the hands $x_{t-1}^{H}$ and body $x_{t-1}^{B}$ to independently learn their denoising distributions. We describe the conditional discriminator D, which depends on the time step and audio control signal, as accepting the noise sequence at step $t-1$, the noise step $t$, and the conditional feature $c$. In our adversarial learning strategy, fake samples from distribution $p_{\theta}(x_{t-1}|x_{t})$  will compete with real samples from distribution $q(x_{t-1}|x_{t})$. D is a 7-layer MLP network, composed of Linear layers, SELU activation, and GroupNorm layers. All models are trained using the AdamW optimizer with a fixed learning rate L. We employ EMA decay for the optimizer during training. For more training settings, please see the implementation details in the experiment. Finally, our final training objective of proposed method is:
\begin{equation}
\begin{aligned}
    \underset{\theta}{\min} \underset{D_{adv}}{\max}  \mathbb{E}_{q(x_{0})q(x_{t-1}|x_{0})q(x_{t}|x_{t-1})}[-log(D_{adv}(x_{t-1},c,t))] \\
    +[-log(1-D_{adv}(\hat{x}_{t-1},c,t))] + \lambda_{recon}(\mathcal{L}_{simple}+\mathcal{L}_{foot})\\
    + \lambda_{AFD}\frac{(1-\beta_{t})||\hat{x}_{t-1}-x_{t-1}||^{2}}{\beta_{t}},
\end{aligned}
\end{equation}
where $\lambda_{recon}$ represents the reconstruction weight of the denoiser, and $\lambda_{AFD}$ represents the weight of the regularization term.
\begin{table*}
\caption{Objective Metrics. The FGD evaluation model is trained across the entire training set, with lower evaluation values indicating closer adherence to the original motion distribution. The 'Inf.time' metric is computed by statistically inferring the time taken to generate each frame (measured in milliseconds); lower values signify faster generation speeds.}
\label{tab:obj}
\centering 
\resizebox{1.0\linewidth}{!}{
\begin{tabular}{lccccccccccc}
\toprule[1.5pt]
\multirow{1}{*}{} & \multicolumn{5}{c}{ZeroEGG} & \multicolumn{5}{c}{BEAT} \\ 
\cline{2-11}
& \multicolumn{1}{c}{FGD$\downarrow$} & \multicolumn{1}{c}{BA$\uparrow$} & \multicolumn{1}{c}{DIV$\uparrow$} & \multicolumn{1}{c}{Inf. time$\downarrow$} & \multicolumn{1}{c}{steps} & \multicolumn{1}{c}{FGD$\downarrow$} & \multicolumn{1}{c}{BA$\uparrow$} & \multicolumn{1}{c}{DIV$\uparrow$} & \multicolumn{1}{c}{Inf. time$\downarrow$} & \multicolumn{1}{c}{steps} \\ 
\cline{2-7} 
\hline
DSG~\cite{yang2023diffusestylegesture}            & 15.67     & 0.81 &0.63     & 6.01  & 1000     & 113.7 &\textbf{0.89} & 0.71 & 5.92 &1000 \\ 
FreeTalker~\cite{yang2024freetalker}    &17.12 &0.72 &\textbf{0.84}  &5.5 &1000 &147.2 &0.84 &\textbf{0.77} &5.5 &1000\\
DiffGesture(re-train)~\cite{zhu2023taming}  &25.7 &0.58 &- &4.72  &1000  &382.6 &0.61 & - & 5.02  & 1000 \\
\textbf{Ours}           & \textbf{12.57}    &\textbf{0.87}  &  0.76   &\textbf{0.29}  &10  &\textbf{98.6} &0.88 &0.74 & \textbf{0.38}  &10\\ 
\hline
\textbf{Ours(DDPM)}    & 11.95    &0.92  &  0.85   &11.9  &1000  &91.7 &0.91 &0.76 & 11.94  &1000\\ 
\hline
Trimodal(re-train)~\cite{yoon2022genea} &22.4 &0.72 &0.80 &- &- &222.3 &0.77 &0.68 &- &-\\
HA2G(re-train)~\cite{liu2022audio}&18.8 &0.81 &0.73 &- &- &156.8 &0.82 &0.70 &- &-\\
CAMN~\cite{liu2022beat} &15.52 &0.83 &0.77 &- &- &258.4 &0.73 &0.61 &- &-  \\
\hline
\end{tabular}
}
\end{table*}

\section{Experiments}
In this section, we evaluated the effectiveness of our proposed method in concurrent gesture generation. To verify the generalizability of the approach, we also supplemented the experiments with an extension in the music-driven dance domain. We compared the performance and computational efficiency of our approach with other state-of-the-art techniques in these domains. By applying our method to real-time tasks, we showcased its robust functionality and promising potential. We highly recommend readers to consult the supplementary video material to gain further insights into the qualitative outcomes.\\
\textbf{Data and Representation.} Our experiments employed three distinct high-quality 3D motion capture datasets: BEATs~\cite{liu2022beat}, ZeroEGGs~\cite{ghorbani2023zeroeggs}, and AIST++~\cite{li2021ai}. The first two were used for concurrent gesture generation, and the latter for real-time dance synthesis. Each dataset contained style labels. For the BEATs dataset, we selected the corresponding English audio data for model training, while the entirety of the training sets was used for the other datasets. Given that our experiments considered the motion of all joints (body and fingers), the dimension of our motion representation significantly increased. This, coupled with physical challenges such as foot sliding and global displacement jitter, made the training more challenging when compared to half-body methods~\cite{yoon2020speech,ao2022rhythmic,yoon2022genea,liu2022audio,zhu2023taming}, but concurrently improved the visual outcomes. We uniformly downsampled the original data to 20FPS for concurrent gesture training and to 30FPS for dance data. Since ZeroEGGs is not in Vicon standard and cannot be directly used to drive the SMPLX standard character model, we converted it to the Vicon standard using MotionBuilder, ensuring consistency with BEATs. Regarding the motion representation for each individual joint, we employed a 6D rotation representation $r\in\mathbb{R}^{6}$, 3D keypoints $l\in\mathbb{R}^{3}$, angular velocity $l\in\mathbb{R}^{6}$, and linear velocity $l\in\mathbb{R}^{3}$ to describe motion information, while also incorporating gaze direction $z\in\mathbb{R}^{3}$ to depict head movements during speech. All gesture data was segmented into 80-frame training clips, and dance data into 150-frame segments. To facilitate the generation of smooth, long-frame motions, we appended a tenth of the seed poses to each segment. Given that all datasets provided discrete style labels, we uniformly transformed them into one-hot encodings.\\
\textbf{Implementation Details}. Our real-time diffusion generation framework is end-to-end, implemented exclusively using Pytorch. Additionally, we developed a script for Blender~\cite{community2018blender} that can accept generated motion sequences in real-time, driving character feedback for users and achieving state-of-the-art visual effects in human-computer interaction. In our implementation, using a V100 GPU, we generated 80 frames of concurrent gestures in 0.4 seconds and 150 frames of dance motions in 0.88 seconds. For a fair comparison with contemporary methods, we conducted experiments on a single V100 GPU. By default, we utilized a single A100 GPU for model training. For the concurrent gesture model, we trained the generator and discriminator for 80 hours using a batch size of 128 and learning rates of 3e-5 and 1.25e-4, respectively. The dance model was trained for 48 hours with a batch size of 128 and learning rates of 5e-5 and 1.5e-4. We set the default diffusion steps to 20 (although similar results can be achieved with 10 steps, offering faster speed) for evaluating all metrics. Moreover, in CFG~\cite{ho2022classifier}, we set the conditional weight to 3.5. In the denoising model, we set the weights of the KL loss and Geo Loss to 0.5 and 10, respectively. These hyperparameters were found to yield the best empirical results.



\begin{table*}
    \caption{Comparative results with contemporary accelerated diffusion methods are presented. Ensuring fairness, all methods employ 10 - step denoising.}
    \label{tab:speed}
    \centering
    \resizebox{1.0\linewidth}{!}{
        \begin{tabular}{lrrrrrrrrrr}
            \toprule[1.5pt]
            \multirow{1}{*}{} & \multicolumn{5}{c}{ZeroEGG} & \multicolumn{5}{c}{BEAT} \\ 
            \cmidrule(lr){2 - 6}\cmidrule(lr){7 - 11}
            & \multicolumn{1}{c}{FGD$\downarrow$} & \multicolumn{1}{c}{BA$\uparrow$} & \multicolumn{1}{c}{DIV$\uparrow$} & \multicolumn{1}{c}{Inf. time$\downarrow$} & \multicolumn{1}{c}{steps} & \multicolumn{1}{c}{FGD$\downarrow$} & \multicolumn{1}{c}{BA$\uparrow$} & \multicolumn{1}{c}{DIV$\uparrow$} & \multicolumn{1}{c}{Inf. time$\downarrow$} & \multicolumn{1}{c}{steps} \\ 
            \midrule
            DPM - Solver - 1(DDIM)~\cite{lu2022dpm} & 21.7 & 0.76 & 0.60 & 0.27 & 10 & 203.7 & 0.71 & 0.66 & 0.17 & 10 \\
            DPM - Solver - 2~\cite{lu2022dpm} & 19.92 & 0.70 & 0.54 & 0.39 & 10 & 148.0 & 0.68 & 0.71 & 0.47 & 10 \\ 
            \midrule
            Naive DDGAN~\cite{wang2022diffusion}  & 44.7 & 0.47 & - & 0.22 & 10 & 334.2 & - & - & 0.25 & 10\\
            Naive SIDDMs~\cite{xu2023semi} & 49.2 & 0.49 & - & 0.27 & 10 & 309.0 & - & - & 0.28 & 10\\ 
            \midrule
            \textbf{Ours} & \textbf{12.57} & \textbf{0.87} & 0.76 & \textbf{0.29} & 10 & \textbf{98.6} & 0.88 & 0.74 & \textbf{0.38} & 10\\ 
            \bottomrule
        \end{tabular}
    }
\end{table*}

\subsection{Comparison with Existing Methods} 
We first evaluated the efficiency and generation quality of our method and compared it with other contemporary open-source diffusion and non-diffusion generation techniques. For the ZeroEGGs dataset~\cite{ghorbani2023zeroeggs}, we assessed various styles, including happiness, sadness, anger, aging, neutral, fatigue, and relaxation. For the BEATs dataset~\cite{liu2022beat}, we selected four speaker sequences as the data used in this study. All comparison methods employed WavLM for extracting audio features. We split the original dataset into training, validation, and test sets with proportions of 0.8, 0.1, and 0.1, respectively, and trained on the entire training set. During the training process, we also train our DDPM implementation with the same structure. For the BEATs dataset, we retrained DSG, DiffGesture~\cite{zhu2023taming} and Trimodal~\cite{yoon2022genea} using open-source code.

\noindent\textbf{Quantitative Comparison.} It is well-known that evaluating a model's generative capability based solely on a limited number of generated examples is challenging; therefore, we introduce several metrics. (i) Fréchet Gesture Distance (FGD)\cite{yoon2020speech} calculates the distance between the latent feature distributions of generated gestures and actual gestures, thereby assessing gesture quality. Lower FGD values indicate higher motion quality. (ii) We compute the number of frames generated per second during the inference stage to demonstrate our outstanding generation efficiency. (iii) We also compared the Beats Alignment (BA) and Diversity (DIV) on the BEAT dataset using the same evaluation method as described in the original paper. Moreover, as shown in Table~\ref{tab:obj}, it is noteworthy that our semi-implicit structure effectively maintains alignment with the audio control task, preserving the optimal FGD score while achieving nearly a 15-fold speed improvement.
\subsection{Comparison with Contemporary Acceleration Strategies} In the Table~\ref{tab:speed}, we compared our method with other acceleration strategies tailored for diffusion-based generation methods. Specifically, our experimental results were compared with strategies employing DPM-Solver and the original DDGAN~\cite{wang2022diffusion} and SIDDMs~\cite{xu2023semi} acceleration. The experiments showed that for the DPM-Solver strategy, its first-order Taylor expansion form corresponds to the well-known DDIM sampling strategy. Accelerating sampling merely by reducing the sampling step size often leads to inaccurate approximations for complex multi-modal distributions, resulting in a sharp decline in quality. Due to the presence of second-order derivatives, the second-order Taylor expansion requires calling the denoising function twice at midpoint positions during sampling, thus reducing generation speed while providing limited improvement in generation quality. Higher-order Taylor expansions require more frequent calls to the denoising function, which contradicts the primary goal of acceleration.

Moreover, we also compared the original DDGAN with SIDDMs acceleration strategy. Specifically, we trained an unconditional discriminator for DiffGes on the BEAT dataset, eliminating explicit geometric loss (which is different from our method), and compared it with our approach. The results (as shown in the table) indicate that the original configuration of implicit and semi-implicit strategies perform extremely poorly in terms of generated global gesture quality. This is because, unlike images, human representations typically have more stringent geometric conditions that require more specific constraints.

\subsection{Ablation Studies} 
In this section, we investigated the impact of diffusion steps, reconstruction loss weight, and auxiliary forward loss on model performance. All ablation studies were conducted on the ZeroEGGs dataset, and the results are presented in Table~\ref{tab:abl}.\\\textbf{Sampling Step Ablation.} In this experiment, we study the effect of different sampling steps on model performance. Specifically, we train models with the same structure using 1, 5, 10, 20, 30, and 50 steps, respectively. The final results, as shown in Table~\ref{tab:abl}, indicate that the FGD value stabilizes after 10 steps, but an increase in the number of steps also leads to slower speed. When the number of steps is 1, our structure reverts to a traditional GAN model, and the generated gesture quality declines sharply. \\
\textbf{Reconstruction Loss Impact.} When the reconstruction loss weight is set to 0, the gesture quality degrades significantly. The introduction of explicit reconstruction loss notably improves the gesture quality. Based on empirical evidence, we set the weight for the experiment to 1, 10, 100 and observe that the weight magnitude does not affect the FGD metric and generation quality. The results without reconstruction loss can be seen in the accompanying video.\\
\textbf{Forward Noise Constraint Impact.} The term "w/o" indicates that we eliminate the forward loss and train the model solely based on adversarial learning. It is worth noting that pure adversarial training, although capable of helping the model learn the marginal distribution, still results in a large gap between the posterior sampling outcome and the forward process noise at the same time step due to the complexity of the large-step distribution, leading to a significant decline in FGD.\\
\begin{table}
\resizebox{1.0\linewidth}{!}{
\begin{tabular}{lccccccccc}
\toprule[1.5pt]
&\multicolumn{3}{c|}{Steps} & \multicolumn{2}{c|}{Recon loss}  & \multicolumn{2}{c}{AFD loss}\\ \cline{2-8}
                  & num      & FGD$\downarrow$     & Inf. time$\downarrow$    & weight     & FGD $\downarrow$      &   & FGD$\downarrow$  \\ \cline{2-7} \hline
     & 1         & 85.44     & 0.03          & 0  & 93.10    & w/o & 26.9\\ 
     & 5         & 20.91     & 0.23          & 1  & 12.74    & w/ & \textbf{12.32} \\
     & 10        & 12.57     & \textbf{0.29}          & 10 & \textbf{12.32}     \\
     & 20        & 12.32     & 0.4          & 100 & 12.38    \\
     & 30        & 12.71     & 0.64          & -   & -        \\
     & 50        & \textbf{12.02}     & 0.99         & -   & -        \\  
     \hline
\end{tabular}
}
\caption{Ablation Experiments. We find that when the diffusion steps reach 10, the FGD metric stabilizes, while it degrades sharply when reduced to a traditional GAN network with only one step. The absence of body reconstruction loss severely impacts the generated quality, but the weight of this constraint has little influence on model learning. AFD can explicitly constrain the difference between forward noise and noise sampled from the denoising distribution at the same time step, further enhancing the quality of the generated gesture sequences.}
\label{tab:abl}
\vspace{-10pt}
\end{table}

\section{Conclusion}
In this paper, we address the speed limitations of diffusion models in generating collaborative speech gestures and explore the challenge of modeling complex denoising distributions across multiple sampling steps. Unlike previous text-to-motion approaches, we introduce implicit marginal constraints based on audio control signals and explicit auxiliary forward diffusion regularization. This improves the model's ability to fit audio control, perform denoising with larger step sizes, and reduce the number of steps, resulting in faster inference. Additionally, we decouple body and finger movements and independently model finger and body distributions. Our approach generates more diverse finger movements while maintaining stability compared to non-physically-based training methods, enhancing the user's viewing experience. These groundbreaking attempts have significantly accelerated DIDiffGes while maintaining high-fidelity generation results, providing new insights for future real-time simultaneous gesture generation tasks. In the future, we will incorporate some gesture datasets based on reconstruction~\cite{cheng2023bopr,liang2024ropetp} to enhance the performance of DIDiffGesture. 

\bigskip

\bibliography{aaai25}

\end{document}